\documentclass[preprint,12pt]{elsarticle}

\usepackage{amssymb}

\usepackage{graphicx,color}
\usepackage[usenames,dvipsnames]{xcolor}
\usepackage[section]{placeins}
\usepackage[normalem]{ulem}

\journal{Physics Letters B}

\begin{document}

\begin{frontmatter}
\title{Zero energy scattering calculation in Euclidean space}
\author{J.~Carbonell}
\address{Institut de Physique Nucl\'eaire,
Universit\'e Paris-Sud, IN2P3-CNRS, 91406 Orsay Cedex, France}
\author{V.A.~Karmanov}
\address{Lebedev Physical Institute, Leninsky Prospekt 53,
119991 Moscow, Russia}

\begin{abstract}
We show that the Bethe-Salpeter equation for the scattering amplitude in the limit of zero incident energy 
can be transformed into a purely Euclidean form, as it is the case for the bound states.
The decoupling between Euclidean and Minkowski amplitudes is  only possible for zero energy scattering observables   and allows determining the 
scattering length from the Euclidean Bethe-Salpeter amplitude.
Such a possibility strongly simplifies the numerical solution of the Bethe-Salpeter equation
and suggests  an alternative way to compute the scattering length in Lattice Euclidean calculations without using the Luscher formalism.
The derivations contained in this work were performed for scalar particles and one-boson exchange kernel. They can be generalized  to the fermion case and more involved interactions.
\end{abstract}
\end{frontmatter}

\section{Introduction} \label{intro}

Bethe-Salpeter (BS) equation \cite{bs} is an efficient tool for study the relativistic systems in an explicitly covariant framework. 
One of the important properties of this formalism is that the formal object it deals with -- the BS amplitude -- has a Quantum Field Theoretical
definition as a vacuum expectation value of the T-ordered product of  Heisenberg field operators  \cite{GML_PR84_51}. For instance, the two-body BS amplitude 
in a scalar theory $\phi$   reads 
\begin{equation}\label{Phi_QFT}
\Phi(x_1,x_2;p) =  \langle  0 \mid T\left\{ {\phi}(x_1) \phi(x_2) \right\} \mid p \rangle 
\end{equation}
where $\mid p\rangle$ is the state  state vector  with total momentum $p=k_1+k_2$.
The BS amplitude $\Phi$  obeys an integral equation which uses to be written in momentum space.
For that purpose, one first remarks that translational invariance imposes $\Phi$ to have the  form 
\[ \Phi(x_1,x_2; p)=\frac{1}{(2\pi)^{3/2}} \; {\Phi}(x,p) \;e^{-ip \cdot(x_1+x_2)/2}, \]
with $x=x_1-x_2$ and where ${\Phi}(x,p)$ is called the reduced amplitude. 
Its Fourier transform  $\Phi(k,p)$  defines the momentum space BS amplitude usually computed:
\[  \Phi(x,p)= \int  d^4x  \;\Phi(k,p)\; e^{-ik\cdot x}. \]
The amplitude $\Phi(k,p)$ is related to the two-body scattering amplitude $F(k,p)$ of the process $k_{1s}+k_{2s}\to k_1+k_2$ as follows
$$
\Phi(k,p)= S_1(k,p)S_2(k,p) F(k,p),
$$
where $S_{1,2}$  are the free one-body propagators and $2k=k_1-k_2$.
Written in terms of $F$, the inhomogeneous BS equation for two spinless particles in Minkowski space reads:
\begin{equation}\label{BSE}
F(k;p)=K(k,k_s)- i\int\frac{d^4k'}{(2\pi)^4} \frac{K(k,k') F(k';p)}
{\left[\left(\frac{p}{2}+k'\right)^2-m^2+i\epsilon\right]  \left[\left(\frac{p}{2}-k'\right)^2-m^2+i\epsilon\right]} \ ,
\end{equation}
where the interaction kernel $K$ is given by a set of the irreducible Feynman graphs. 
The different steps of this derivation are detailed in \cite{IZ,Greiner_QED}.

For the bound state case, the inhomogeneous term drops out and the $p$-dependence of the off-mass shell amplitude $F(k;p)$ in the center of mass (c.m.) frame appears
only via $p^2=M^2$ where $M$ is the total mass of the system. 
The notation $F(k_0,\vec{k})$ is then currently used.

The off-mass shell scattering amplitude, $F(k;p)$  depends also on a parameter $2k_s=k_{1s}-k_{2s}$ -- the incident relative momentum --  which  in general is not related to $p$.
For the results we are interested in (half off-mass shell quantities in c.m. frame) the $p$-dependence of the amplitude is however related to $k_s$ by $p^2=M^2=4\varepsilon^2_{k_s}$ 
and will be replaced in the notation by $k_s$. We will write hereafter:  $F(k;p)\equiv F(k;k_s)\equiv F(k_0,\vec{k};k_s)$.
Taken on the mass shell $k_0=0$, $k=k_s$,   the value $F(k_0=0,k=k_s)$  determines the phase shifts.
 
The original BS equation (\ref{BSE}) was formulated in Minkowski space. Its numerical solution  -- made very difficult  by the singularities of the free propagators, the interaction kernel and the amplitude itself --
was not even tried until very recently \cite{Kusaka,bs1-2,Sauli_JPG08,fsv-2012,fsv-2014}.

For the bound state case the BS equation can be transformed, by the Wick rotation \cite{WICK_54} of the integration contour of the variable $k_0$, in an Euclidean form over the variable $k_4=-ik_0$. 
The Euclidean BS amplitude  $F_E$ is related to the Minkowski one $F_M$ by 
\[ F_E(k_4,\vec{k})=F_M(ik_4,\vec{k}) \]
This change in the metric removes all the singularities for real values of the arguments  and solving the corresponding Euclidean equation becomes an easy  numerical task. 
It can be shown, at least for some simple interaction kernels \cite{WICK_54,bs1-2},  that the Wick rotation let invariant the total mass $M$ and represents thus an efficient way to compute  the binding energies.  
However the off-mass shell amplitude $F(k_0,k)$  in the Minkowski space  remains mandatory to calculate the electromagnetic elastic  form factors \cite{ck-trento}.

In contrast to the bound state, the scattering states BS equation  cannot be in general transformed into an equation for  the Euclidean amplitude alone.
As we will see later, when rotating the integration contour, some singularities are crossed and one must take into account  the residues in these poles  \cite{tjon}.
These residues are proportional to the BS amplitude in Minkowski space at the particular value   $k_0$=$\varepsilon_{k}-\varepsilon_{k_s}$, with
$\varepsilon_{k}$=$\sqrt{m^2+k^2}$ and correspondingly for $\varepsilon_{k_s}$. 
Applying the Wick rotation one obtains thus a system of coupled equations for the two amplitudes 
$F_E(k_4,k)$ and $F_M(k_0=\varepsilon_{k_s}-\varepsilon_{k},k)$. 
This system of equations was derived in \cite{tjon, bs_long}. 
Solving this system of equations is a simpler task than solving the initial BS equation in Minkowski space but does not provide the full off-shell solution of the scattering problem.
Since the on-shell condition $k_0=0$ is equivalent to $k_4=0$,  the  Minkowski and Euclidean on-mass shell amplitudes  coincide with each other  $F_E(k_4=0,k_s)=F_M(k_0=0,k_s)$
and  the Euclidean amplitude $F_E(k_0,k)$ obtained by solving the coupled equations is enough to provide the phase shifts. 

The difficulties  for solving directly the scattering state BS equation in Minkowski space  were recently overcome. 
A first method is  based on  the Nakanishi integral \cite{nak63} representation of  the BS amplitude  and solving the resulting equation for the  Nakanishi weight function. 
It was successfully applied to  bound states \cite{Kusaka,bs1-2,Sauli_JPG08}. 
In Ref. \cite{fsv-2012} a scattering equation for the Nakanishi weight function was also derived.  It was  solved for  zero incident momentum in \cite{fsv-2015}. 

A second method  taking analytically into account all the singularities  was developed and proved its efficiency for both the bound and, especially, the scattering states \cite{CK_PLB_2013,bs_long}. 
Corresponding  bound-to-scattering state transition   electromagnetic form factor was found in \cite{tff_long}.

The main aim of this work is to show that in the limit of zero incident momentum ($k_s\to 0$) the  Euclidean BS amplitude $F_E(k_4,k)$ decouples from the Minkowski
part and can be thus obtained by solving a single integral equation in Euclidean metric.
Like in the case of the bound state problem, the solution of the Euclidean equation determining $F_E(k_4,k)$  can be obtained with high precision and  low numerical cost. 
Once $F_E(k_4,k)$  is determined,  the scattering length is given by $a_0=-F_E(0,0)/m$.
The possibility, here illustrated, to obtain a scattering length from an Euclidean solution can have interesting applications in Lattice QCD calculations.

It is worth mentioning that an  Euclidean  formulation of the relativistic quantum mechanics
has been recently developed in \cite{Polyzou}.

The paper is organized as follows.
In Sec. \ref{sec1} we derive the Euclidean BS equation for the spinless  scattering amplitude at zero incident momentum $k_s=0$.  The numerical results for the 
scattering length (for the ladder kernel)  and comparison with the results found in Ref. \cite{fsv-2015} are given in Sec. \ref{num}.  Sec. \ref{concl} contains discussion and concluding remarks.

\section{Deriving Euclidean equation}\label{sec1}

We will derive in this section  the integral equation satisfied by the off-shell Euclidean amplitude $F_E(k_4,k)$, corresponding to the zero incident momentum $k_s=0$. 

One possible way to proceed would be to consider  the coupled Euclidean-Minkowski system of equations derived in Ref. \cite{bs_long} (appendix C) and study their limit when $k_s\to0$.
However we prefer to present here  an independent  and  self consistent derivation of this equation.

To this aim we will first consider the singularities of the four-dimensional equation (\ref{BSE}),
find an appropriate integration contour ensuring the Wick rotation for non-zero $k_s$ and  finally take the limit $k_s\to 0$. 

In this study  we will consider the one-boson exchange kernel:
\begin{equation}\label{obe}
K(k,k')=-\frac{16\pi m^2\alpha}{(k-k')^2-\mu^2+i\epsilon},
\end{equation}
with \mbox{$\alpha=g^2/(16\pi m^2)$}  is the dimensionless coupling constant.  

Calculations are performed in the c.m.  frame, defined by $\vec{p}=0$.  In this reference system 
one has, by definition of the incident momentum $k_s$, \mbox{$p_0=2\varepsilon_{k_s}=2\sqrt{m^2+k_s^2}$}.
The on-shell conditions $k_1^2=k_2^2=m^2$, in terms of variables $k_0,\,k$, are reduced to $k_0=0$, $k=k_s$.

The pole singularities associated with the propagators in (\ref{BSE}) are given by 
\begin{eqnarray}
{k'}^{(1)}_0(k,k_s)&=&\phantom{-}\varepsilon_{k_s}+\varepsilon_{k'} - i\epsilon = +a_+  -   i\epsilon            \cr
{k'}^{(2)}_0(k,k_s)&=&\phantom{-}\varepsilon_{k_s}-\varepsilon_{k'}+ i\epsilon  = -a_-    +  i\epsilon      \cr
{k'}^{(3)}_0(k,k_s)&=&-\varepsilon_{k_s}+\varepsilon_{k'}- i\epsilon  = +a_- - i\epsilon      \cr
{k'}^{(4)}_0(k,k_s)&=&-\varepsilon_{k_s}-\varepsilon_{k'}+ i\epsilon  = - a_+ + i\epsilon   \label{Poles}
\end{eqnarray}
with
\begin{equation}\label{a+-}
a_{\pm}(k',k_s) =\varepsilon_{k'}   \pm \varepsilon_{k_s} 
\end{equation}

\begin{figure}[htbp]
\centering\includegraphics[width=7.cm]{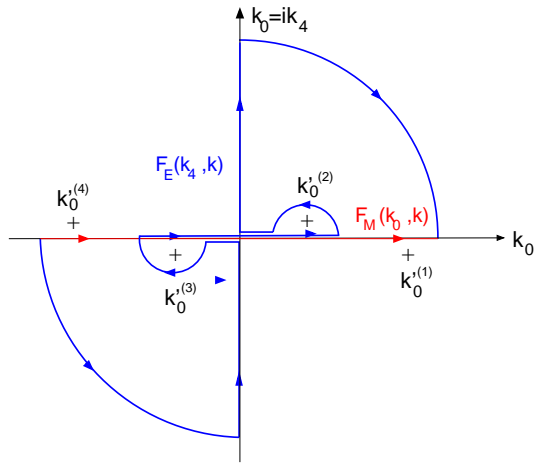}
\caption{Singularities of the propagators for scattering state and the integration contour after rotation in the complex plane $k_0$, if $k'<k_s$.}\label{fig1a}
\end{figure}

In the case $k'<k_s$, their positions in the complex plane $k_0$ are shown in Fig. \ref{fig1a}. 
When the integration contour is rotated,  the   singularities 
${k'}^{(2)}_0$ and ${k'}^{(3)}_0$ are crossed and the corresponding residues of the integrand at these poles should be taken into account.

Equation (\ref{BSE})  is then transformed into:
\begin{equation}\label{BSEc}
F^E(k_4,\vec{k};\vec{k}_s)=V^B(k_4,\vec{k};\vec{k}_s)+ \int\frac{d^4k'}{(2\pi)^4} \frac{V(k_4,\vec{k'};k'_4,\vec{k'}) F^E(k'_4,\vec{k'};\vec{k}_s)} {({k'_4}^2 +a_-^2)({k'_4}^2+a_+^2)} + S(k_4,k,k_s)
\end{equation}
where the Wick-rotated interaction  kernel (\ref{obe})  reads
\begin{equation}\label{V}
V(k_4,\vec{k};k_4',\vec{k'})=\frac{16\pi m^2\alpha}{(k_4-k'_4)^2+(\vec{k}-\vec{k'})^2+\mu^2},
\end{equation}
and the Born term is given by
\begin{equation}\label{Vinh}
V^B(k_4,\vec{k};\vec{k}_s)=V(k_4,\vec{k};k_4'=0,\vec{k'}=\vec{k}_s)
\end{equation}
The remaining term $S(k_4,k,k_s)$ denotes the contribution of the two  poles ${k'}^{(2)}_0$ and ${k'}^{(3)}_0$ shown in Fig. \ref{fig1a}. 

The residual contribution -- existing only if $k'<k_s$ -- is  the sum of two terms $S=S_1+S_2$:
 $S_1$ is the residue at the pole  ${k'}_0^{(2)}=-a_- + i\epsilon$   and $S_2$ the one from ${k'}_0^{(3)}=+a_-  - i\epsilon$ (multiplied by $2i\pi$).

The first term $S_1$ is given by:
\footnotesize
\begin{equation}\label{sing1}
S_1(k_0)=\frac{\pi g^2}{4(2\pi)^4} \int_{k'<k_s}d^3k' \frac{\tilde{F}_M(k',z')}{\varepsilon_{k_s} \varepsilon_{k'}[-a_- +i\epsilon]
\left[-a_- + \sqrt{(\vec{k'}-\vec{k})^2+\mu^2} -k_0\right]  \left[-a_- -  \sqrt{(\vec{k'}-\vec{k})^2+\mu^2} -k_0 +i\epsilon\right]}  
\end{equation}
\normalsize
where $\tilde{F}_M(k',z')$ is the particular value of the Minkowski amplitude 
\[  \tilde{F}_M(k',z')=  F(k'_0=\varepsilon_{k_s}- \varepsilon_{k'},k')\]
and  {$d^3k'={k'}^2dk'd\phi dz'$. 

$S_2$  is given by a simple substitution $S_2(k_0)=S_1(-k_0)$. Notice that the sum $S_1+S_2$ is symmetric relative to $k_0\to - k_0$, as it should be. 

Setting $k_0=ik_4$ we finally  obtain for $S=S_1+S_2$:
\begin{small}
\begin{equation}\label{sing}
S=\frac{g^2 \pi}{(2\pi)^4}  \int_{k'<k_s}d^3k'\frac{\tilde{F}_M(k',z')} {2\varepsilon_{k'} \varepsilon_{k_s}( a_- -i\epsilon)}   
\frac{\Bigl[k_4^2-\Bigl(a_- -\sqrt{(\vec{k'}-\vec{k})^2+\mu^2}\,\Bigr) \Bigl(a_- +\sqrt{(\vec{k'}-\vec{k})^2+\mu^2}\,\Bigr)\Bigr]}
{\Bigl[k_4^2+\Bigl(a_- -  \sqrt{(\vec{k'}-\vec{k})^2+\mu^2}\,\Bigr)^2\Bigr] \Bigl[k_4^2+\Bigl(a_- +\sqrt{(\vec{k'}-\vec{k})^2+\mu^2}\,\Bigr)^2\Bigr]}   
\end{equation}
\end{small}

This expression, as well as Eq. (\ref{BSEc}) contains integrable singularities which make difficult the numerical solution. 
However, it is worth noticing that this problem disappears in the limit  $k_s=0$.
Indeed, the integrand in (\ref{sing}) contains the singular factor
\begin{equation}\label{prop}
\frac{1}{ a_- -i\epsilon}=\frac{1}{\varepsilon_{k'}-\varepsilon_{k_s}-i\epsilon}=PV\frac{1}{\varepsilon_{k'}-\varepsilon_{k_s}}+i\pi\delta(\varepsilon_{k'}-\varepsilon_{k_s})=PV\frac{{\varepsilon_{k'}+\varepsilon_{k_s}}}{{k'}^2-{k_s}^2}+i\pi\delta(\varepsilon_{k'}-\varepsilon_{k_s})
\end{equation}
In the limit $k_s\to 0$ the singularity of the principal value is cancelled by the factor ${k'}^2$ from the integration volume
and in the shrinking limits $0<k'<k_s$ the integral (\ref{sing}) tends to zero. 
Therefore the contributions of two singularities shown  in Fig. \ref{fig1a} disappear: $S\to 0$ when $k_s\to 0$.

The last term $S$ in Eq. (\ref{BSEc}) can be omitted and the four-dimensional BS equation for the Euclidean amplitude in the zero energy limit takes the form
\begin{equation}\label{BSEc0}
F^E(k_4,\vec{k};0)=V^B(k_4,\vec{k};0)+ \int\frac{d^4k'}{(2\pi)^4} \frac{V(k_4,\vec{k'};k'_4,\vec{k'}) } {({k'_4}^2 +a_-^2)({k'}^2_4+a_+^2)} \; F^E(k'_4,\vec{k'};0)
\end{equation}

This is a self-consistent integral equation for determining $F^E(k_4,\vec{k};\vec{k}_s=0)$ without any coupling to the Minkowski amplitude $F$.
The possibility of such a decoupling, is the vanishing of the $S$ term in Eq.  (\ref{BSEc}) and it is valid only in the limit $k_s\to 0$.
 
 \bigskip
To obtain the scattering length, it is of practical interest to make the partial wave decomposition and isolate the S-wave contribution. 
Following the convention  \cite{IZ}:
\begin{equation}\label{fpw1}
F^E(k_4,\vec{k};0)=16\pi\sum_{L=0}^{\infty}(2L+1) F_L(k_4,k)P_L(z),
\end{equation}
where $P_L(z)$ is the Legendre polynomial and $F_L$ is the partial  amplitude depending only on $k_4$ and  $k=\mid\vec{k}\mid$.
Since we are hereafter  restricted to the case $k_s=0$,  we omit  the argument $k_s$ as well as the Euclidean label. 

In the case of S-wave,  Eq. (\ref{BSEc0}) reduces to the following two-dimensional integral equation:
\begin{equation}\label{eq1}
F_0(k_4,k)=V_0^B(k_4,k) +\int_0^{\infty}{k'}^2dk'\int_0^{\infty}dk'_4\frac{V_0(k_4,k;k'_4,k')}{({k'_4}^2+{a'_-}^2)({k'_4}^2+ {a'_+}^2)}\;F_0(k'_4,k'),
\end{equation}
where $ a'_\pm=\sqrt{m^2+{k'}^2}\pm m$ and
\begin{eqnarray}\label{VB}
V_0^B(k_4,k)&=&\frac{\alpha m^2}{k_4^2+\mu^2+k^2}      \label{V0B}  \\ 
V_0(k_4,k;k'_4,k')  &=&\frac{m^2\alpha}{\pi^2 kk'}\log\frac{[k_4^2+{k'_4}^2+\mu^2+(k+k')^2]^2-4k_4^2{k'_4}^2}  {[k_4^2+{k'_4}^2+\mu^2+(k-k')^2]^2-4k_4^2 
{k'_4}^2}  \label{Vs}
\end{eqnarray}
Notice that in  Eq.   (\ref{eq1})   we have limited the $k_4$ integration to the half-interval $0<k_4<\infty$.

The factor $({k'_4}^2+{a'_-}^2)$ in the denominator of (\ref{eq1})  vanishes at $k'_4=0,\;k'=0$.  In practical calculations this (integrable) singularity can be removed by the replacement of variable
$k'_4=a'_-y$ and use the relation $a'_+a'_-={k'}^2$. Then Eq. (\ref{eq1}) obtains the form:
\begin{equation}\label{eq2}
F_0(k_4,k)=V_0^B(k_4,k)
    +\int_0^{\infty}a'_+ dk'\int_0^{\infty} dy\frac{V_0(k_4,k;ya'_-,k')}{(y^2+1)(y^2{a'_-}^2+ {a'_+}^2)}\;F_0(ya'_-,k')
\end{equation}
This is the final non-singular equation for the Euclidean off-shell S-wave amplitude $F_0(k_4,k)$ corresponding to the zero incident momentum $k_s=0$. As mentioned, taken on mass shell, $F_0(k_4,k)$ determines the scattering length:

\begin{equation}\label{eq3}
a_0=-\frac{1}{m}F_0(k_4=0,k=0)
\end{equation}

In Ref. \cite{bs_long} we have obtained a system of two coupled equations -- Eqs. (C1) and (C8) --  relating the partial wave amplitudes in Euclidean and Minkowski spaces at arbitrary incident momentum $k_s$.  
One can show that in the limit $k_s\to 0$,  equation (C1)   reproduces Eq. (\ref{eq2}). 
Note however that Eq. (C1) contains a few extra terms in comparison to (\ref{eq2}) which in the limit $k_s\to 0$ either disappear or cancel each other.  
This cancellation is not straightforward and,  to be proved, requires some analytical transformations.

\bigskip
The interest in deriving equation (\ref{eq1})    is not only a significant simplification in the way to compute the scattering length in the BS framework 
but in the fact that this fundamental quantity can be obtained from the Euclidean solution. 
Indeed,  the Euclidean BS amplitude (\ref{Phi_QFT}) has been here calculated by solving the BS equation in the ladder approximation and with a simple one boson exchange kernel but it is actually accessible in the full Quantum Field Theory solution provided by the Lattice approach.
For instance, it has been the basic ingredient of the HAL-QCD Lattice collaboration  to obtain the first {\it ab initio} Nucleon-Nucleon (NN) potential  from QCD
 \cite{Ishii:2006ec,Kurth:2013} but  never been used to extract the scattering observables in the way we propose.
 Notice however than in Ref.  \cite{Kurth:2013}, the  NN potential  computed this way was  inserted  in the Schrodinger equation whose solution  provided the scattering length. 
This approach, involving  several uncertainties and approximations related to the inverse scattering problem, is in contrast with the direct  method we suggest and which
according to (\ref{eq3}) requires  only to compute the momentum space BS amplitud at the $k_0=0=0$ values.
 See e.g. Ref. \cite{Aoki:2012tk} and references therein for the history and technical details of HAL-QCD method.

The possibility to obtain scattering amplitudes from Euclidean correlator in infinite space 
is forbidden by the so called Maiani-Testa no-go theorem \cite{Maiani_Testa_PLB245_1990}. This result
is in agreement with the impossibility discussed above to obtain scattering observable without a coupling
to the Minkowski amplitudes, but does not apply to the zero scattering energy where the phase shifts vanish, as was already pointed out in \cite{ Maiani_Testa_PLB245_1990,Testa2}.

On the other hand M. Luscher and collaborators \cite{ML_CMP104_86,LL_CMP219_03}
circumvented this problem taking benefit from what is in fact a limitation of any Lattice approach -- its finite volume $V$ -- and
proposed a method based on the $V$-dependence  of the two-particle energies confined in a box with periodic boundary conditions.
This approach has been very successful in computing scattering length and phase shifts of several hadronic systems from {\it ab initio} 
QCD \cite{Fukugita:1994ve,Aoki:1999pt,Beane:2003da,Beane:2005rj,Feng:2009ij,Metivet}. One can find a more complete reference list in the reviews 
\cite{Beane:2010em,Detmold:2015jda}. 

The use of  equation  Eq. (\ref{eq3}) constitutes  an alternative method to compute the scattering length  in the Lattice calculations,
directly from the Fourier transform  of the Euclidean  BS amplitude defined in  (\ref{Phi_QFT}).
A similar suggestion was considered in \cite{Testa2} in connection with Luscher formalism  and for finite volume calculation.
In the  framework of BS equation, such a possibility is justified by the existence of Eq. (\ref{eq1}) allowing to compute the Euclidean BS amplitude in a purely Euclidean formalism  as it is the case in the Lattice approach.

The numerical results presented in the next section validate this approach in the simple case of the one boson exchange model
where the results can be found by independent methods.

\section{Numerical results}\label{num}

The numerical solutions of Eq.  (\ref{eq2}) with the kernel (\ref{Vs}) have been obtained by spline expansion of the Euclidean  amplitude $F_0(k_4,k)$ and solving the corresponding linear system (see Appendix A of Ref. \cite{bs_long} for details).

The scattering  lengths $a_0$ are extracted by computing the value at the origin of the Euclidean amplitude (\ref{eq3}).
Their values are in  full agreement with the results of our previous work \cite{CK_PLB_2013}  and from those of Ref. \cite{fsv-2015}, both obtained using different and independent methods. 
Some deviations in the case $\mu=0.15$, noticed in Table 1 of \cite{fsv-2015},  were due to  inaccuracies in \cite{CK_PLB_2013}  and
have been corrected in benefit of \cite{fsv-2015}.
This lack of precision  in the  $\mu=0.15$ results of   Ref. \cite{CK_PLB_2013}  was due to an unadapted choice of the grid parameters for large values of the coupling constant $\alpha$. 
In view of the agreement with the preceding results (Tab. 1 of \cite{CK_PLB_2013}  and Tabs. 1-3 of \cite{fsv-2015}), it would be redundant to repeat here the numerical values. 
However, we give below the Euclidean off-shell amplitude $F_0(k_4,k)$ determining by Eq. (\ref{eq3}) the scattering length.

\vspace{0.5cm}
\begin{figure}[htbp]
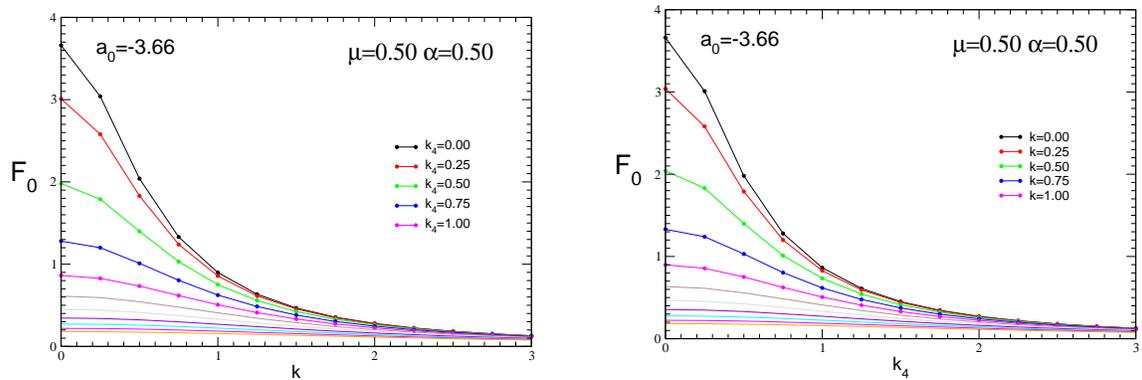

\centering\includegraphics[width=7.cm]{F_k_mu_0.50_a_0.50_L_5_N_20.eps}\hspace{0.9cm}
\centering\includegraphics[width=7.cm]{F_k4_mu_0.50_a_0.50_L_5_N_20.eps}
\caption{Euclidean scattering amplitude $F_0(k_4,k)$ as a function of $k$ for different values of  $k_4$ (left) and as a function of  $k_4$  for different values of $k$ (right). They correspond to $m=1$, $\mu=0.50$ and $\alpha=0.50$.
The  scattering length value is given by $a_0=-F(0,0)=-3.66$.}\label{F_mu_0.50_a_0.50}   
\end{figure}

Figure \ref{F_mu_0.50_a_0.50} displays the amplitude $F_0(k_4,k)$ as a function of $k$ at fixed values of $k_4$ (left panel) 
and as a function of $k_4$  at fixed values of $k$ (right panel) for the parameters $m=1$, $\mu=0.50$ and $\alpha=0.50$.
The scattering length $a_0=-3.66$ is directly readable in both panels.
For these parameters  the two-body system has no bound state and the amplitudes are monotonic function in both arguments.

The same amplitude is shown in  Figure \ref{F_mu_0.15_a_2.50} for $\mu=0.15$ and $\alpha=2.50$.
For these parameter values the two-body system has two bound states with  the second one having very small binding energy. 
This results in a  large and positive value of the scattering length $a_0=+12.3$.
The amplitude has consequently a richer structure in both variables than in Fig. \ref{F_mu_0.50_a_0.50} and  requires hence a finer grid to be properly described.

\vspace{0.5cm}
\begin{figure}[htbp]
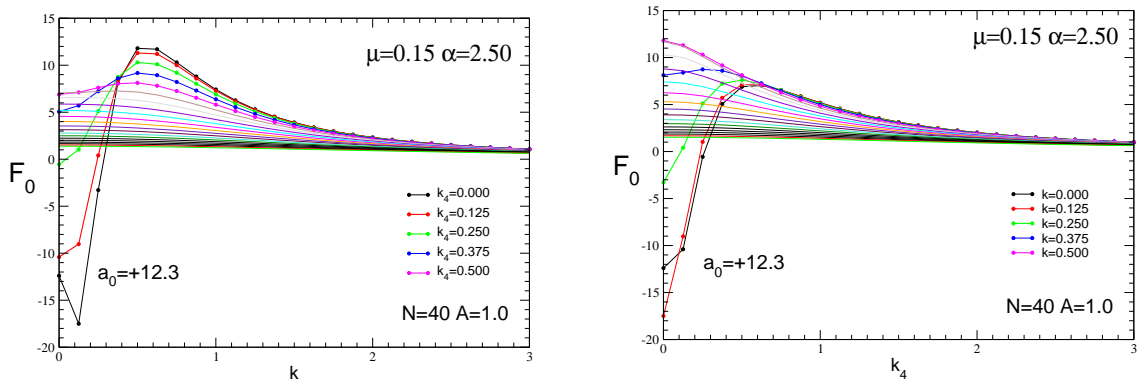

\centering\includegraphics[width=7.cm]{F_k_mu_0.15_a_2.50_L_5_N_40.eps}\hspace{0.9cm}
\centering\includegraphics[width=7.cm]{F_k4_mu_0.15_a_2.50_L_5_N_40.eps}
\caption{The same than in Fig. \ref{F_mu_0.50_a_0.50} with the parameters $\mu=0.15$ and $\alpha=0.50$. The  scattering length value is given by $a_0=-F(0,0)=+12.3$}\label{F_mu_0.15_a_2.50}   
\end{figure}

\section{Conclusion}\label{concl}

We have shown that in the limit of zero incident energy the Bethe-Salpeter  scattering amplitude  
can be obtained by solving a purely Euclidean equation, as it was the case for the bound states.

The decoupling between Euclidean and Minkowski Bethe-Salpeter amplitudes is  only possible for zero energy scattering observables   and allows determining the 
scattering length from the Euclidean amplitude alone.

These results have been stablished and tested in the framework of the  Bethe-Salpeter  equation with a scalar ladder one-boson exchange kernel,  
where the scattering lengths can be obtained by independent methods.
They can be generalized to the fermion case, since scalar and  fermion propagators have the same singularities. 
The validity of the Wick rotation was already stablished with more elaborated kernels, like the cross-ladder one \cite{bs1-2}.

Our results suggest the possibility  to directly extract the scattering length in Lattice  calculations by  computing  the Euclidean
Bethe-Salpeter amplitude  (\ref{Phi_QFT}) in momentum space,  without using the Luscher formalism.


\end{document}